\documentclass[iop]{emulateapj}
\usepackage{natbib}

\def\psr{PSR~J2222}
\def\PSR{PSR~J2222$-$0137}
\def\Tref#1{Table~\ref{tab:#1}}
\def\Fref#1{Figure~\ref{fig:#1}}
\def\Sref#1{Section~\ref{sec:#1}}

\newcommand{\Teff}{\ensuremath{T_{\rm eff}}}
\newcommand{\til}{\raisebox{-0.7ex}{\ensuremath{\tilde{\;}}}}

\newcommand{\expnt}[2]{\ensuremath{#1 \times 10^{#2}}}   

\renewcommand{\apjl}{ApJL}

\slugcomment{ApJ, in press}
\shorttitle{The Coolest Known White Dwarf?}
\shortauthors{Kaplan et al.}

\begin{document}

\title{A 1.05\,$M_\odot$ Companion to PSR~J2222$-$0137: The Coolest
  Known White Dwarf?}

\author{David L.~Kaplan\altaffilmark{1}, 
Jason Boyles\altaffilmark{2,3},
Bart H.~Dunlap\altaffilmark{4},
Shriharsh P.~Tendulkar\altaffilmark{5},
Adam T.~Deller\altaffilmark{6},
Scott M.~Ransom\altaffilmark{7},
Maura A.~McLaughlin\altaffilmark{2,8}, 
\&\ Duncan R.~Lorimer\altaffilmark{2,8}
}

\altaffiltext{1}{Department of Physics, University of
Wisconsin-Milwaukee, 1900 E. Kenwood Boulevard, Milwaukee, WI 53211,
USA; kaplan@uwm.edu}
\altaffiltext{2}{Department of Physics and Astronomy, West Virginia University, White Hall, Box 6315, Morgantown, West Virginia 26506-6315, USA}
\altaffiltext{3}{Physics and Astronomy Department, Western Kentucky
  University, 1906 College Heights Boulevard \#11077, Bowling Green,
  Kentucky 42101-1077, USA}
\altaffiltext{4}{Department of Physics and Astronomy, University of North Carolina, Chapel Hill, NC, 27599-3255, USA}
\altaffiltext{5}{Space Radiation Laboratory, California Institute of Technology, 1200 E California Blvd, MC 249-17, Pasadena, CA 91125, USA}
\altaffiltext{6}{ASTRON, P.O.~Box 2, 7990 AA Dwingeloo, The
  Netherlands}
\altaffiltext{7}{National Radio Astronomy Observatory, 520 Edgemont Road, Charlottesville, Virginia 22903-2475, USA}
\altaffiltext{8}{Also adjunct at National Radio Astronomy Observatory, Green Bank, WV 24944, USA}


\begin{abstract}
The recycled pulsar \PSR\ is one of the closest known neutron stars,
with a parallax distance of $267_{-0.9}^{+1.2}\,$pc and an edge-on
orbit.  We measure the Shapiro delay in the system through pulsar
timing with the Green Bank Telescope, deriving a low pulsar mass ($1.20\pm0.14\,M_\odot$) and a high
companion mass ($1.05\pm0.06\,M_\odot$) consistent with either a
low-mass neutron star or a high-mass white dwarf.  We can largely reject the
neutron star hypothesis on the basis of the system's extremely low
eccentricity ($\expnt{3}{-4}$)---too low to have been the product of
two supernovae under normal circumstances.  However, despite deep optical and near-infrared
searches with SOAR and the Keck telescopes we have not discovered the optical counterpart of the
system.  This is consistent with the white dwarf hypothesis only if
the effective temperature is $<3,000\,$K, a limit that is robust to
distance, mass, and atmosphere uncertainties.  This would make the
companion to \PSR\ one of the coolest white dwarfs ever observed.  For
the implied age to be  consistent with the age of the Milky Way
requires the white dwarf to have already crystallized and entered the
faster Debye-cooling regime.
\end{abstract}

\section{Introduction}

\object[PSR J2222-0137]{PSR~J2222$-$0137} (hereafter \psr) is a 33\,ms radio pulsar discovered
in the Green Bank Telescope (GBT) 350 MHz drift-scan pulsar survey \citep{blr+13}. With a
dispersion measure of 3.27\,pc\,$\rm cm^{-3}$, it appeared to be one
of the closest pulsars to the Earth.  Further observations showed
\psr\ was in a binary system with an orbital period of 2.45\,days and a
minimum companion mass of about 1\,$M_\odot$. This sort of system
straddles the line between potential companion types.  It
could be a double-neutron star (DNS), of which there are only roughly
12 and whose study is crucial to understanding the formation of
sources of kHz gravitational waves \citep[e.g.,][]{kkl03} and testing
general relativity \citep[e.g.,][]{stairs10}.  Or, it could be a
pulsar with a massive white dwarf companion---a so-called
``intermediate-mass binary pulsar'' (IMBP)---that descended from a
binary with a more massive companion than in  traditional systems
with pulsars and low-mass white dwarfs  
\citep*{vdh04,tvdhs00,tlk11,tlk12}.  IMBP systems are rare, with fewer
than 20 known, and massive white dwarfs are themselves rare, 
with fewer than 8\% of the white dwarfs (WDs) from optical surveys
having masses above $0.9\,M_\odot$ \citep*{gbr11}.  
Understanding the formation and evolution of IMBP systems provides
a crucial piece in our understanding of binary evolution and pulsar
recycling, and helps delineate evolutionary paths between low-mass
NSs and high-mass white dwarfs \citep{tauris11}.


\citet{dbl+13} used very long baseline interferometry astrometry to measure the parallax of
\psr\ with exquisite precision.  They find a distance of
$267_{-0.9}^{+1.2}\,$pc (it is the second closest binary pulsar system
and one of the closest NSs of any type).  The astrometric
data also suggested an edge-on orbit, opening up the possibility of a
measurement of the Shapiro delay \citep{shapiro64}, which gives two
post-Keplerian \citep{lk12} parameters for the system and hence
determines the component masses \citep[e.g.,][]{dpr+10}.  Here we
present the detailed timing analysis of the \psr\ system, including
the measurement of the Shapiro delay and the determination of the
masses (\Sref{gbt}).  We then present deep optical and near-infrared
searches for the companion to \psr\ (\Sref{oir}), which we use to
constrain models of its formation and evolution (\Sref{disc}).  We
find that the system almost certainly must be an IMBP system, but that
we do not detect the companion, constraining it to be one of the
coolest white dwarfs ever observed.  Unlike some sources where
temperature inferences are highly dependent on white dwarf model
atmospheres \citep[e.g.,][]{ggh+04}, this measurement is robust, given
the small uncertainties on the mass and (especially) distance.  We
conclude in \Sref{conc}.



\section{Observations and Analysis}
\subsection{Radio Observations}
\label{sec:gbt}
Radio observations of \psr\ to measure the Shapiro delay occurred in
the last week of 2011 May with the 100\,m Robert C.~Byrd GBT\footnote{The Robert C. Byrd Green Bank Telescope (GBT)
  is operated by the National Radio Astronomy Observatory which is a
  facility of the U.S. National Science Foundation operated under
  cooperative agreement by Associated Universities, Inc.}.  We had a
6\,hr observation taken around superior conjunction of the binary
system augmented by five 2\,hr observations at each of the other five
Shapiro extrema, all using the Green Bank Ultimate Pulsar Processing
Instrument (GUPPI; \citealt{drd+08}).  The 800\,MHz of bandwidth
centered at 1500\,MHz in two orthogonal polarizations was separated
into 512 Nyquist-sampled frequency channels of width 1.5625\,MHz via a
polyphase filter bank.  These channels, sampled at 8-bits, provided
full polarization information and an effective time resolution of
0.64\,$\mu$s.  Each channel was coherently dedispersed at the nominal
dispersion measure (DM) of the pulsar (3.27761\,pc\,cm$^{-3}$ at the time, although we
later refined this measurement).  Each observing session was broken
into 30-minute observations of \psr\
separated by 60\,s calibration scans of the extragalactic radio source
3C~190.  The calibration scans were taken in the same mode as the
pulsar observations, but also included a 25\,Hz noise diode inserted
into the receiver.

\begin{figure}[t]
\centerline{\includegraphics[width=0.4\textwidth,angle=270]{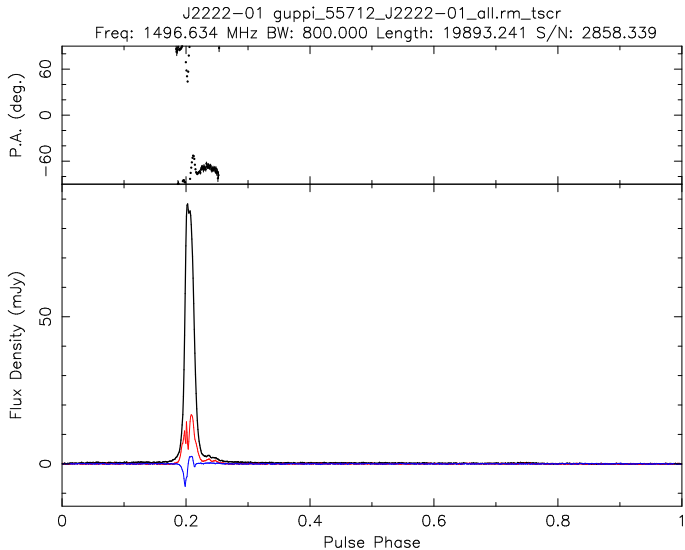}}
\caption{Pulse profile of \psr\ from the GUPPI observation covering
  conjunction.  In the lower panel, black is the total intensity, red
  is linear polarization, and blue is circular polarization (Stokes $V$).  The
  position angle of the linear polarization is given in the upper
  panel. As is the case with most MSPs,
the polarization position angle variations do not permit a  rotating
vector model fit, so we cannot constrain the emission geometry.
\label{fig:profile}
}
\end{figure}

Data reduction was performed using the \texttt{PSRCHIVE} package
\citep*{hvsm04}.  Flux calibration used the on- and off-source scans of
3C~190.  This was followed by removal of  radio
frequency interference  by the  {\tt psrzap} utility.  The
calibrated pulse profile determined from the long observation covering
conjunction is given in \Fref{profile}. 
The data were aligned in time
using the best ephemeris (below), divided into 16 frequency channels,
and re-fit for dispersion measure and rotation measure using a
bootstrap error analysis.  We found that the period-averaged flux density varied
by a factor of a few over the course of long observations due to
scintillation, with an average of 1--2\,mJy at 1500\,MHz.
Individual times-of-arrival (TOAs) were measured from the folded total-intensity profiles using the
frequency domain algorithm in \texttt{PSRCHIVE} 
\citep{taylor92}. A template was created by fitting
three Gaussians to the summed pulse profile.  From these
Gaussian components, we created  a noise-free template with the
phase of the fundamental component in the frequency domain rotated to
zero.  The observations were divided into 2\,minute segments,
with
one TOA measured for each segment.  Note that since interstellar
scintillation caused the flux to vary considerably, there was a
proportional change in the TOA precision that varied over the
data set.

These data were combined with previous data taken for the discovery
observations of \psr\ \citep{blr+13} to produce a timing model.  We
used the ``DD'' model \citep{dd85,dd86} in \texttt{TEMPO},\footnote{\url{http://tempo.sourceforge.net/}.} which incorporates the
Shapiro delay.  The astrometric data for this model were taken from
\citet{dbl+13}, and we used the DE421 JPL ephemeris \citep*{fwb09}.
Timing fits with no Shapiro delay were statistically unacceptable,
with an rms residual of $9.3\,\mu$s ($\chi^2=4539.4$ for 931
degrees-of-freedom), and a clear Shapiro delay signature was obvious
in the residuals (\Fref{sha}).  With the Shapiro delay included in the
fit the rms
residual was 4.2\,$\mu$s ($\chi^2=930$ for 929 degrees-of-freedom),
with no obvious remaining structure in the residuals (varying the
astrometric parameters within the uncertainties from \citealt{dbl+13}
changed the timing results by $\ll 1\,\sigma$).  The Shapiro delay
determines the inclination of the orbit and the companion mass; this
is then combined with the binary mass function to determine the
pulsar's mass.  Due to the combination of several different and much
less precise observing modes from earlier monitoring with the
high-precision Shapiro delay campaign, we estimated the timing
parameters with a bootstrap error analysis.  We give the full timing
results, with 1-$\sigma$ error estimates from the bootstrap analysis,
in \Tref{par}.

\begin{figure}[t]
\centerline{\includegraphics[width=0.5\textwidth]{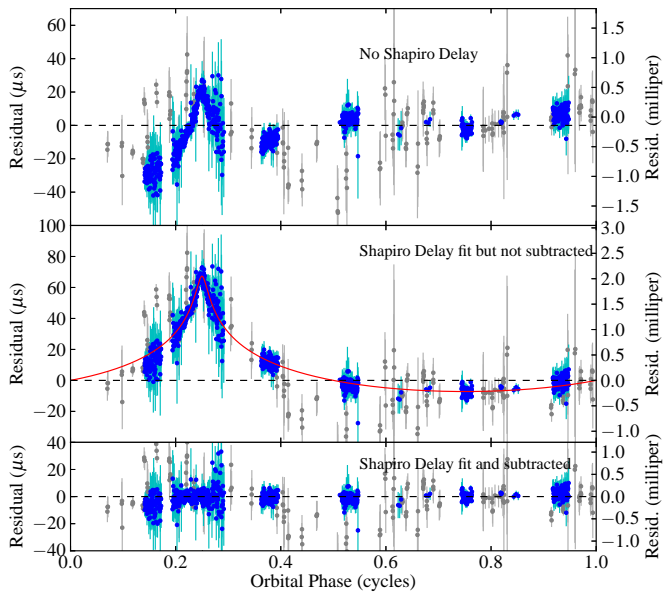}}
\caption{\label{fig:sha} Timing residuals for \psr, using the new data
  from this paper (blue: MJD 55,600--55,921) and older data
  (gray), as a function of orbital phase (true anomaly plus longitude
  of periastron).  Top: residuals computed from the best-fit model
  without Shapiro delay (the rms residual is $9.3\,\mu$s).  Middle:
  residuals computed including Shapiro delay.  The red curve is the
  best-fit Shapiro delay profile.  Bottom: residuals computed relative
  to the best-fit model including Shapiro delay (the rms residual is
  $4.2\,\mu$s).  Conjunction is at a phase of 0.25.
In all panels the left axis shows the
residuals in $\mu$s, while the right axis shows the residuals in
milliperiods.  Note the different $y$-axis scales.}
\end{figure}

{Our data consist of high-quality coherently dedispersed data
  from an intensive 1 week campaign and a few other epochs.  The
  remainder of the data were both less precise and less uniform, with a
  wider range of observation frequency and instrumental setup.  This
  makes it difficult (if not impossible) to robustly constrain
  long-term secular changes like periastron precession ($\dot \omega$;
  \citealt{lk12}).  Nonetheless, we tried a fit with $\dot \omega$
  fixed to the value predicted by general relativity ($\approx
  0\fd08\,{\rm yr}^{-1}$).  The resulting fit was good, with the
  rms decreasing to 3.8\,$\mu$s.  The pulsar and companion masses each
  increased by about 1$\sigma$ compared to the values in \Tref{par}.
  Given the small eccentricity and inhomogeneous data set with large
  gaps we do not believe that fitting for $\dot \omega$  is viable
  at this time, but encourage further long-term monitoring of this
  system to establish its secular behavior.  }

\begin{deluxetable}{l c}
\tabletypesize{\footnotesize}
\tablewidth{0pt}
\tablecaption{Fitted and Derived Timing Parameters for \PSR.\label{tab:par}}
\tablehead{
\colhead{Parameters} & \colhead{Value}}
\startdata
\multicolumn{2}{c}{Timing parameters} \\
\hline
Spin period (s)\dotfill & 0.032817859053065(3)\\
Period derivative (s\,s$^{-1}$)\dotfill & $5.865(7)\times 10^{-20}$\\
Dispersion measure (pc\,cm$^{-3}$)\dotfill & 3.2842(6)\\
Rotation measure (rad\,m$^{-2}$) \dotfill & +2.6(1) \\
Reference epoch (MJD)\dotfill & 55743\\
Right ascension\tablenotemark{b} (J2000)\dotfill & 22:22:05.969101(1)\\
Declination\tablenotemark{b} (J2000)\dotfill & $-01$:37:15.72441(4)\\
R.A.\ proper motion\tablenotemark{b} (mas\,$\rm yr^{-1}$) \dotfill &  44.73(4)\\
DEC proper motion\tablenotemark{b} (mas\,$\rm yr^{-1}$) \dotfill & $-$5.68(6) \\
Parallax\tablenotemark{b} (mas) \dotfill &   $3.742^{+0.013}_{-0.016}$  \\
Position epoch\tablenotemark{b} (MJD) \dotfill & 55743  \\
Span of timing data (MJD) \dotfill & 55005--55922\\
Number of TOAs\tablenotemark{a} \dotfill & 943\\
rms residual ($\mu$s) \dotfill & 4.2\\
\hline
\multicolumn{2}{c}{Binary parameters\tablenotemark{d}} \\
\hline
Orbital period (days) \dotfill & 2.4457599929(3)\\
Projected semi-major axis (lt-s) \dotfill & 10.8480276(12)\\
Epoch Of periastron (MJD) \dotfill & 55742.13242(0)\\
Orbital eccentricity \dotfill & $3.8086(15)\times 10^{-4}$\\
Longitude of periastron (deg) \dotfill & 119.778(12)\\
Mass function ($M_\odot$) \dotfill & 0.22907971(8) \\
$\sin i$ \dotfill & 0.9985(3) \\
Companion mass ($M_\odot$) \dotfill & 1.05(6) \\
\hline
\multicolumn{2}{c}{Derived parameters} \\
\hline
Distance\tablenotemark{b} (pc) \dotfill & 267.3$^{+1.2}_{-0.9}$\\
Transverse velocity\tablenotemark{b} (km\,$\rm s^{-1}$) \dotfill & 57.1$^{+0.3}_{-0.2}$  \\
Orbital inclination $i$ (deg) \dotfill & 86.8(4) \\
Shklovskii period derivative ($\rm s\,s^{-1}$) \dotfill & $4.33(5) \times 10^{-20}$\\
Intrinsic period derivative\tablenotemark{c} ($\rm s\,s^{-1}$) \dotfill & $1.54(5) \times 10^{-20}$\\
Surface magnetic field\tablenotemark{c} ($10^{9}$ Gauss) \dotfill & 0.719 \\
Spin-down luminosity\tablenotemark{c} ($\rm 10^{31}\,ergs\,s^{-1}$) \dotfill & 1.72 \\
Characteristic age\tablenotemark{c} (Gyr) \dotfill & 33.8 \\
Pulsar mass ($M_\odot$) \dotfill & 1.20(14)\\
Flux density at 1500\,MHz (mJy) \dotfill & 1--2\\
 \enddata
 \tablecomments{Values in parentheses are uncertainties on the last
   digit.  For the timing data derived here, the uncertainties were
   derived from a bootstrap analysis and are quoted at the 1$\sigma$ level.}
 \tablenotetext{a}{During the initial timing observations we calculated
   a TOA every 10 minutes.  During the new observations described here
   we calculated a TOA every 2 minutes.}
 \tablenotetext{b}{Values are from \citet{dbl+13} and were held fixed
   for the timing fit.}
 \tablenotetext{c}{Values are corrected for Shklovskii  effect.}
 \tablenotetext{d}{We used the ``DD'' model \citep{dd85,dd86}.}
 \end{deluxetable}

\subsection{Optical/IR Observations}
\label{sec:oir}
We observed the position of \psr\ at optical and near-infrared
wavelengths, as listed in \Tref{log}.  The deepest Keck
observations used the red side of the Low-Resolution Imaging
Spectrometer (LRIS; \citealt{occ+95}) on the 10\,m Keck~I telescope.
The data were reduced using standard procedures in \texttt{IRAF},
subtracting the bias, dividing by flatfields, and combining the
individual exposures.  The seeing was about $0\farcs8$ in the combined
$R$ image, and $0\farcs7$ in the combined $I$ image.  We computed an
astrometric solution {fitting for a shift and separate scales and
rotations along each axis (i.e., a six-parameter fit)} using 100 non-saturated stars identified from the
Sloan Digital Sky Survey (SDSS) Data Release 10 (DR10;
\citealt{aaa+14}), giving rms residuals of $0\farcs2$ in each
coordinate.  We did photometric calibration relative to SDSS photometry,
identifying 23 well-detected, well-separated, non-saturated stars, and
transforming from the SDSS filter set to Johnson--Cousins using the
appropriate transformation equations\footnote{See
  \url{http://www.sdss.org/dr5/algorithms/sdssUBVRITransform.html\#Lupton2005}.}.
The zero-point uncertainty was $< 0.01\,$mag, although there are
systematic uncertainties coming from our filter transformations.  We
see no object at the position of the pulsar (\Fref{image}); the
closest object is about $2\arcsec$ from the position of the pulsar
(about $10\,\sigma$ away) and appears extended {($R=23.1\pm0.1$ and
statistical position uncertainties of $\pm 0\farcs3$ in each coordinate)}. We determined the
3\,$\sigma$ upper limits using \texttt{sextractor} \citep{ba96} to
determine the magnitude that gave a 0.3\,mag uncertainty (verified
with fake-star tests), which we give in \Tref{log}.

\begin{figure*}[t]
\plotone{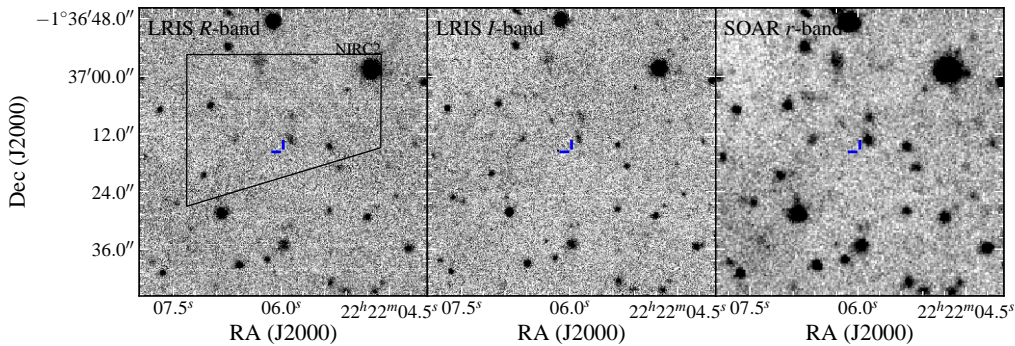}
\caption{Optical images  of the field of \psr: LRIS $R$-band (left),
  LRIS $I$-band (middle), and SOAR $r$-band (right).  The
  position of \psr\ is indicated with the ticks at the center, which
  begin $0\farcs5$ from the pulsar (larger than the position
  uncertainty of the pulsar combined with the astrometric uncertainty
  of the image).  North is up, east to the left, and the image is
  $1\arcmin$ in size.  On the $R$-band image we also indicate the
  field-of-view covered by our NIRC2 image, with the region masked
  apparent in the lower-right.  }
\label{fig:image}
\end{figure*}

We observed \psr\ in $r$-band with the Goodman Spectrograph
\citep*{cca04} on the 4.1\,m Southern Astrophysical Research (SOAR) telescope over two nights in 2013~July.
All exposures were dithered and binned by a factor of two in both
dimensions.  The frames were bias-subtracted and flattened with a dome
flat.  We then used a median of the data (having masked the 
scattered-light halos of three saturated stars) from the second night constructed
without registration to create a sky flat, which we smoothed with a
$20\times 20\,$pixel boxcar filter.  This corrects for larger-scale
brightness variations.  Cosmic rays were interpolated on individual
exposures using the \texttt{lacosmic} routine \citep{vandokkum01}.
The seeing varied considerably over the course of the observations,
going from $1\farcs1$ to $2\arcsec$.  We then shifted each exposure by
an integer number of pixels for registration and summed them.  The
final summed image has an effective seeing of $1\farcs3$ and a total
exposure time of 2.6\,hr.  The photometric zero-point was again
computed relative to the SDSS DR10 data, using 31 stars.  The
astrometric solution was done using six 30\,s exposures through
\url{http://astrometry.net} \citep{lhm+10}.  As with the Keck data, we
see no object at the position of the pulsar (\Fref{image}) and give a 3\,$\sigma$
upper limit in \Tref{log}.

\begin{deluxetable*}{l c c c c c}
\tablewidth{0pt}
\tablecaption{Optical/Near-infrared Observations and Limiting Magnitudes\label{tab:log}}
\tablehead{
\colhead{Instrument} & \colhead{Date} & \colhead{Filter} &
\colhead{Exposure} & \multicolumn{2}{c}{Limiting Magnitude} \\ \cline{5-6}
 & & & & \colhead{Apparent\tablenotemark{a}} & \colhead{Absolute\tablenotemark{b}} \\
 & & & \colhead{(s)} 
}
\startdata
SOAR/Goodman & 2013~Jul~2 & $r$ & $300 + 3\times 600$ & 26.4\tablenotemark{c} & 19.2 \\
SOAR/Goodman & 2013~Jul~3 & $r$ & $18 \times 400$  & & \\
Keck~I/LRIS(red) & 2013~Aug~4 & $R$ & $2\times 300$ & 26.3 & 19.1\\
Keck~I/LRIS(red) & 2013~Aug~4 & $I$ & $2\times 300$ & 26.0 & 18.9\\
Keck~II/NIRC2 & 2013~Oct~12 & $K^\prime$ & $60+5\times 120$ & 21.0 &13.9\\
\enddata
\tablenotetext{a}{3\,$\sigma$ limiting magnitudes at the position of
  the pulsar.}
\tablenotetext{b}{Absolute magnitude limits computed for a distance of
  $267\,$pc and an extinction of $A_V=0.12\,$mag.}
\tablenotetext{c}{The two SOAR observations were combined.}
\end{deluxetable*}

While they were taken through different filters
and with very different instruments/resolutions, we tried combining
the Keck $R$-band and SOAR $r$-band images using \texttt{swarp}
\citep{bmr+02}. We still see no source at the position of the pulsar.
The data are sufficiently different that a limiting flux is difficult
to compute, but it could be as much as 0.3\,mag fainter than the
limits in  \Tref{log}.  

The near-infrared observations come from the NIRC2 camera\footnote{The
  NIRC2 camera can be utilized in three different magnification
  modes. We used the ``wide'' camera with a 40$\arcsec$ square field of
  view.}  on the 10\,m Keck~II telescope, and used the Laser Guide Star
 Adaptive Optics (AO) system \citep{vdblm+06}.  The data were
taken through thin clouds and the AO corrections were not optimal,
resulting in a delivered image quality of $0\farcs2$ FWHM.  The images
were reduced using a custom pipeline implemented with \texttt{python}
and \texttt{pyraf} using dark frames and dome-flats. A sky fringe
frame was created by combining dithered images of multiple targets
with the bright stars masked. We used \texttt{SExtractor}~\citep{ba96}
for the preliminary detection and masking of stars. The fringe frame
was subtracted from the flat-fielded data after being scaled to the
appropriate sky background level. Before coadding the frames, each
frame was corrected for optical distortion using a distortion solution
measured for NIRC2\footnote{See
  http://www2.keck.hawaii.edu/inst/nirc2/forReDoc/post\_observing/dewarp/}. A
faint glare has been visible in the lower right (south-west) corner of
the NIRC2 wide camera images starting in 2009~August. The shape and
amplitude of the glare vary with telescope orientation, resisting
correction through surface fitting or modeling.  Instead we masked the
glare using a triangular region.  There was no independent photometric
calibration that night, and only a single star is visible on the
co-added image.  To determine a photometric zero-point, we used
photometry for that star from the SDSS DR10.  We then employed the
empirical main-sequence color relations from \citet{cis+07}, inferring
the $z-K_s$ color from the observed $g-i$ color (we ignore differences
between $K_s$ and $K^\prime$ filters).  For this star
(SDSS~J222204.76$-$013658.9) we infer a
spectral type of K2.5 and predict $K_s=16.9$.  We expect zero-point uncertainties of
$\pm0.2$\,mag or so based on comparison of the other SDSS colors to
those predicted using \citet{cis+07}.  Again we see no object at the
position of the pulsar, and give 3\,$\sigma$ upper limits in
\Tref{log}.

\section{Discussion}
\label{sec:disc}
\subsection{A Low-mass Neutron Star?}
Since we do not detect the optical counterpart of the companion, the
 first inference is that the companion could be a low-mass
NS.  It would be the lowest mass NS known
\citep{lattimer12,opnsv12,kkdyt13}, although it is only a roughly 2--3
$\sigma$ excursion from the mean of the companions in DNS systems
\citep{opnsv12,kkdyt13}: rare, given the $\approx10\,$DNS systems, but
not impossible.  

In that case, its eccentricity of $\expnt{3.8}{-4}$ would be a factor of $200$ lower
than any other DNS system (\object[PSR J1906+0746]{PSR~J1906+0746}
has the lowest eccentricity of $e=0.085$, although this may be an NS--WD
system; {\citealt{kasian12}}; van Leeuwen et
al.\ 2014, \apj, submitted).  In \Fref{e} we show the eccentricity versus
component masses for all DNS and NS--WD systems with well-determined
masses.  In fact there are three NS--WD systems with higher
eccentricities: \object[PSR J1141-6545]{PSR~J1141$-$6545}, which was
likely not recycled \citep{klm+00}; \object[PSR
  J0337+1715]{PSR~J0337+1715}, which has had its eccentricity increased by
dynamical interactions \citep{rsa+14}; and \object[PSR
  J0621+1002]{PSR~J0621+1002} (likely an IMBP, with the eccentricity
the result of unstable mass transfer; \citealt{pk94,clm+01}).

The normal formation scenario for a DNS involves two core-collapse
supernova explosions, with the eccentricity the result of the second
explosion and its kick, and no final mass-transfer phase to circularize the
orbit \citep[e.g.,][]{tvdh06}.  In contrast, formation via an
electron-capture supernova (ECS; \citealt{mnys80}) could result in a
significantly lower NS mass (\citealt*{spr10}; \citealt{fsk+13}) along with a lower
supernova kick \citep{plp+04,vdh04}.  \psr\ has a low transverse velocity
($58\,{\rm km\,s}^{-1}$), although higher than some systems thought to
be the products of ECSs {(given the age of the system, this velocity
may be more related to motion in the Galactic potential than birth conditions)}.  This may reflect the velocity dispersion of
the progenitor systems.  However, the contrast between \psr\ and other
systems thought to be the results of ECSs (e.g., PSR~J1906+0746 or
\object[PSR J0737-3039]{PSR~J0737$-$3039}; \citealt{fsk+13}) is
extreme, with the ratio of eccentricities above 200 as mentioned
previously.  In a scenario without a kick we can place an upper limit
on the amount of material that could have been ejected by the
explosion to $(M_{\rm PSR}+M_c)e=\expnt{8}{-4}\,M_\odot$
\citep[with $M_{\rm psr}$ the pulsar mass and $M_{\rm c}$ the current
  companion mass; e.g.,][]{bvdh91}.  This is a much tighter bound than in any of
the other systems proposed for this mechanism, and difficult to
reconcile with the change in binding energy needed to collapse to a NS
$\sim G M_c^2/R_c c^2\approx 0.1\,M_\odot$ (with $R_c\approx 15\,$km
for an NS), presumably released as neutrinos
\citep[e.g.,][]{ft14}: this leads to the horizontal line in \Fref{e},
above which all confirmed DNS systems are found.
In order to have a DNS system with such a low eccentricity, we need to
invoke increasingly exotic  (and perhaps implausible) evolutionary
scenarios.  For instance, if the system began as  a hierarchical  triple
\citep{crl+08,rsa+14,tvdh14}, then the inner components could have formed a
standard eccentric DNS system early on.   Later evolution of the outer
member could have led to a circum-binary accretion disk that would
have worked to circularize the inner system, after which the outer
object would have exploded or otherwise been ejected from the system.

\begin{figure}[t]
\plotone{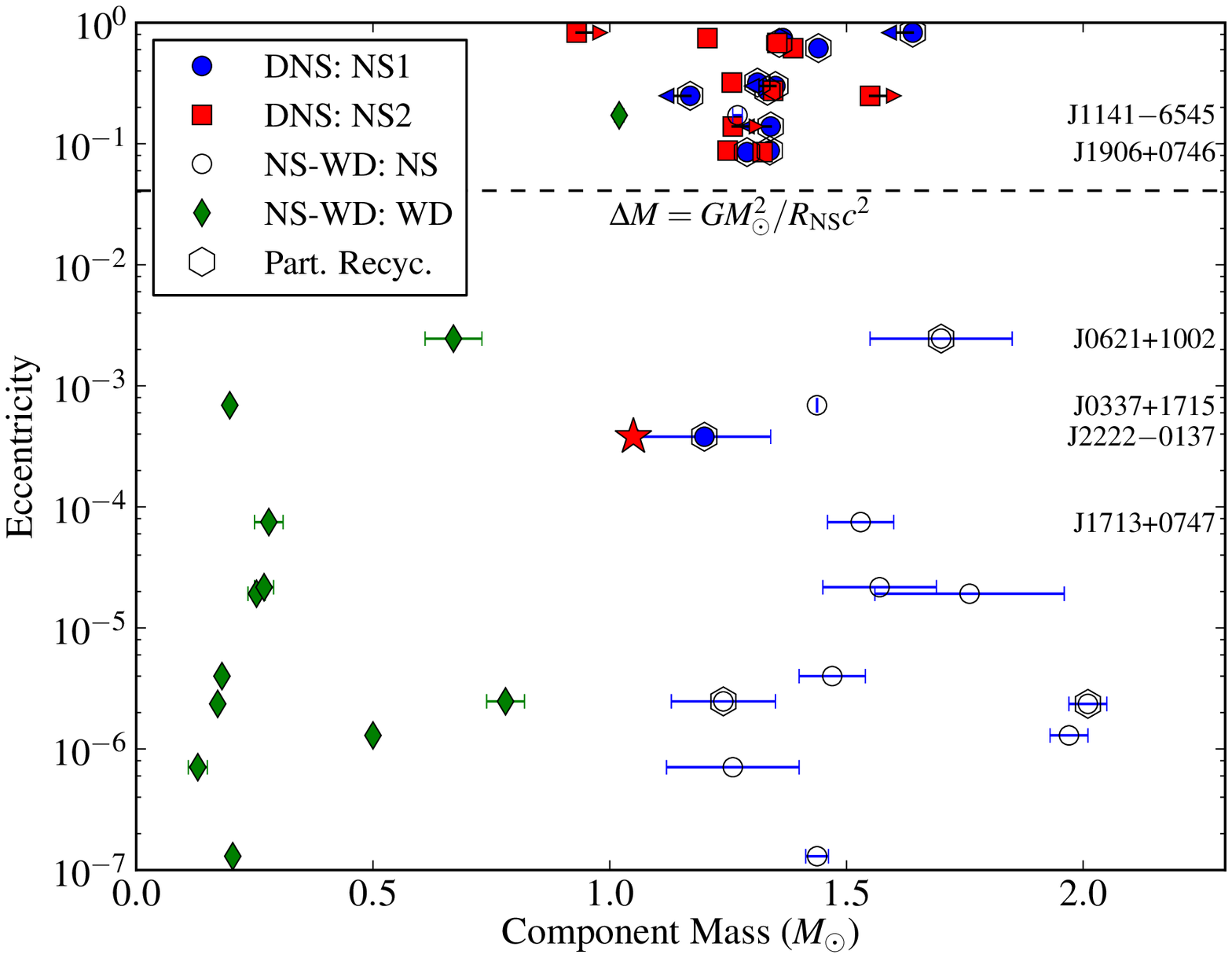}
\caption{Mass vs.\ orbital eccentricity for DNS systems and NS--WD
  systems.  The data are those systems with well-determined (from
  radio timing and optical spectroscopy) masses
  from \citet{fsk+13} and \citet{kkdyt13}, augmented by
  \citet{kkl+09}, \citet{lfrj12},
  \citet{avkk+12}, \citet{afw+13}, and \citet[][inner companion only]{rsa+14}.  For the DNS systems the primary (recycled) NS
  is the blue circle, while the secondary NS is the red square.  For
  the NS--WD systems the NS is an open circle, while the WD is a green
  diamond.  Those systems that are only partially recycled (with spin
  periods between 10\,ms and 0.2\,s) are indicated by hexagons. We
  exclude the NS--WD systems in globular clusters, where the
  eccentricity can be influenced by dynamical encounters.  In
  some cases the nature of the companion (WD versus NS) is not clear,
  such as in PSR~J1906+0746 (van Leeuwen et al.\ 2014) and \psr\ (the red star).  Selected
  systems are labeled.  The horizontal line is given by $e=\Delta
  M_c/(M_c+M_{\rm NS})$ \citep{bvdh91}, with $\Delta M_c=G
  M_\odot^2/R_{\rm NS}c^2$ the minimum change in binding energy needed
  to form an NS \citep{ft14}: all confirmed DNS systems are found above
  this line.
}
\label{fig:e}
\end{figure}


\subsection{An Intermediate-Mass Binary Pulsar?}
The other possible scenario is that the companion could be a massive
WD, making the system an IMBP.
Its orbital eccentricity is somewhat high compared to most low-mass
binary pulsars of similar periods (based on \citealt{phinney92}), but
not nearly as high as a DNS, consistent with an IMBP classification
\citep{clm+01}.  {It falls  in the locus of other CO WDs in the
``Corbet'' (binary period versus spin period) diagram in \citet{tlk12}.
  The pulsar mass is lower than most pulsar--WD binaries, but is
  consistent with the short orbital-period IMBP discussed by
  \citet{fsk+10} which may indicate a similar formation mechanism
  involving a common envelope \citep{tvdh06}.
}

However, as a WD it would be extremely faint: far fainter than any of
the optical companions to IMBPs currently known
\citep{vkbjj05,jcvk+06,pld+13} or indeed any WD companion to a
millisecond pulsar (MSP)
with a similar mass \citep{abw+11}; it is perhaps the faintest WD ever
observed. With the apparent magnitude limits
from \Tref{log}, we can compute absolute magnitude limits in each
band.  We use the distance $267\pm1\,$pc \citep{dbl+13}, and we
estimate the extinction to be $A_V=0.12\,$mag from \citet*{dcllc03}.
In terms of bolometric luminosity the most constraining limit ends up coming from the $R$-band data,
where we limit $M_R>19.1$ (the $r$-band limit of $M_r>19.2$ is very
similar, given slight differences in bolometric correction).  For comparison, the companion to
\object[PSR J1022+1001]{PSR~J1022+1001} with a median companion mass
of $0.85\,M_\odot$ has $M_R\approx 14$ \citep*{lfc96}.  In \Fref{mass}
we plot the absolute magnitude against mass for pulsar+WD systems as
well as select cool WDs with parallax distances: even compared to the
observed truncation of the cooling sequence in old halo globular clusters like
NGC~6397 \citep{rgh+13} or M4 \citep{bsp+09}, the putative companion
is far fainter: at the distance of NGC~6397, our limit of $M_R>19.1$
translates to an apparent magnitude of $R>31.6$, compared to $R\approx
29$, or $M_R\approx 16$ for the coolest WDs seen in
NGC~6397.  Some of the
difference comes from the change in radius: a $1.0\,M_\odot$ WD has a
radius about 65\% of that of a typical $0.6\,M_\odot$ WD, leading to a
1\,mag change in brightness at the same effective temperature.  But
the difference in \Fref{mass} is more like 2.5\,mag, so the companion
to \psr\ must also be cooler than the known {thick disk/}halo WDs.

\begin{figure}
\plotone{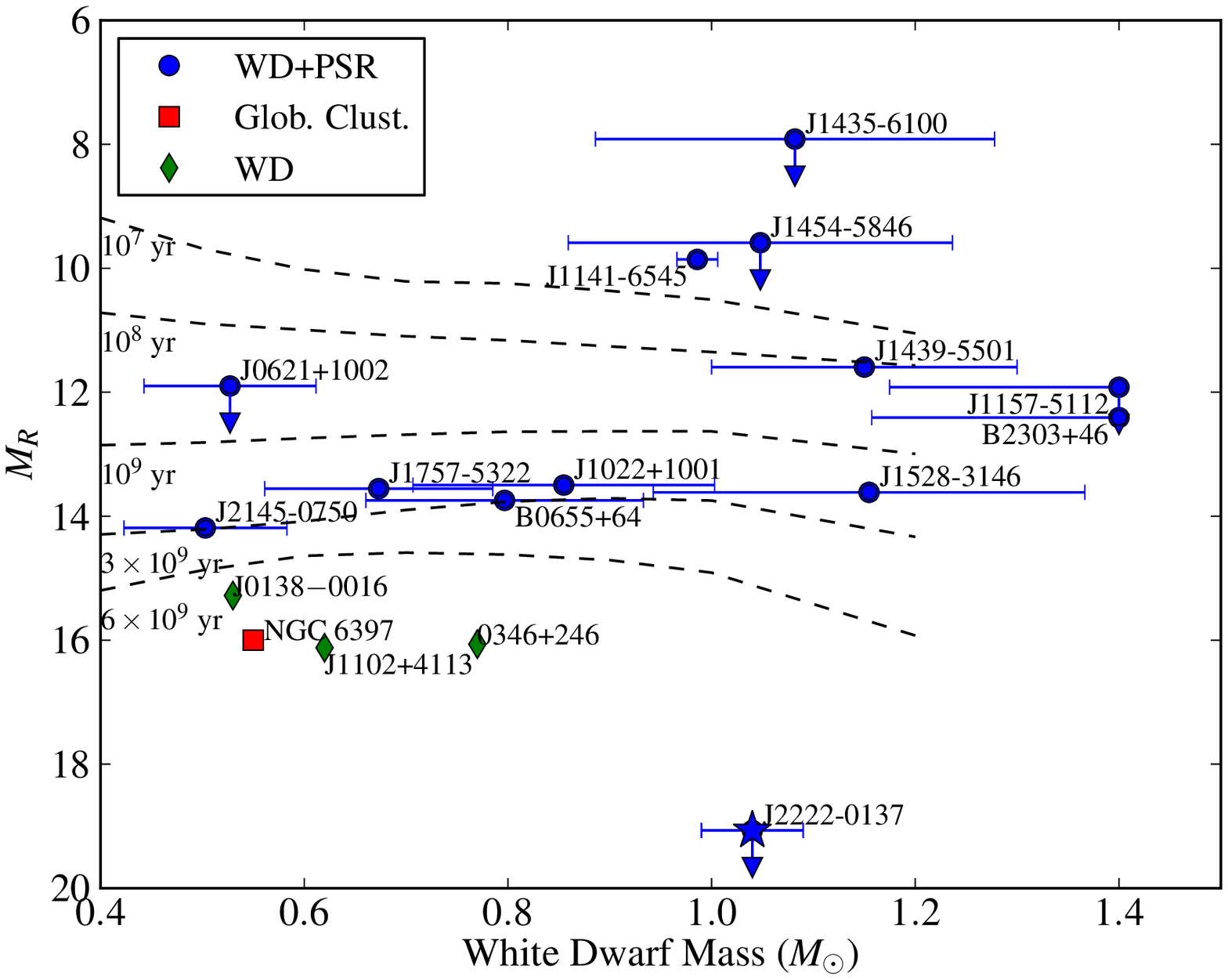}
\caption{Absolute $R$ magnitude plotted against WD mass for massive
  and/or cool WDs.  We show the IMBPs and massive WDs from
  \citet{vkbjj05} and \citet{jcvk+06}, with PSR~J1141$-$6545 updated
  from \citet{abw+11} and PSR~J1439$-$5501 from \citet{pld+13}.  Data
  from bands other than $R$ were converted to $R$ using the photometry
  of \citet*{tbg11} and with extinctions from \citet{dcllc03} computed
  for the distances of the pulsars, {except for updated extinctions
  for PSR~J1439$-$5501 \citep{pld+13} and PSR~B2303+46 \citep{vkk99}}.  All other pulsar data come from
  \citet[][version 1.48]{mhth05}.  {Distances are from \citet{cl02},
  except PSRs~J1022+1001 and J2145$-$0750 (A.~T.~Deller et al., 2014,
  personnal communication), PSR~J1141$-$6545 \citep{obvs02}, and 
  PSR~J1439$-$5501 \citep{pld+13}.}  When the inclination is not
  constrained, the point is at the median value (inclination of
  $60\degr$) but a range is indicated by the error bars, and we allow
  a maximum companion mass of $1.4\,M_\odot$.  We also show the approximate truncation of
  the WD cooling sequence from the halo globular cluster NGC~6397
  (square; \citealt{rab+06,hab+07,rgh+13}), as well as isolated cool halo WDs and
  the eclipsing ultra-cool WD SDSS~J0138$-$0106
  (diamonds; \citealt{ktka12,pgm+12}).  The dashed lines show contours of
  constant age for DA WDs based on \citet{tbg11}, with the ages listed
  at the left.}
\label{fig:mass}
\end{figure}

Beyond the absolute magnitude, which is directly computable from
observable quantities, we can limit the radius/temperature of a
putative WD by using our $R$-band absolute magnitude limit to
constrain the bolometric luminosity.
This is more complicated, as it involves atmosphere
calculations in an uncertain and poorly tested regime, but it should
be reasonably reliable.  We use the synthetic photometry and
evolutionary models from \citet{tbg11} and \citet{bwd+11} for H and He
atmospheres, respectively\footnote{Also see
  \url{http://www.astro.umontreal.ca/{\til}bergeron/CoolingModels/}.}.
For isolated WDs pure He atmospheres can be largely excluded because
of Bondi-Hoyle accretion from the ISM \citep{bergeron01}, and even
small amounts of hydrogen mixed into the helium can cause
near-infrared flux deficiencies like pure hydrogen (see below;
\citealt{bl02}).  However, the binary orbit and MSP wind in this
system could have inhibited such accretion and therefore a He
atmosphere is possible.  In any case a pure He
atmosphere will serve as a limiting case compared to the H models.
These models are used to convert the absolute magnitude limits into
temperature limits, so for simplicity we use the $1.0\,M_\odot$ models
(differences in bolometric corrections as a function of mass are
small, $<0.05\,$mag).

\begin{figure}
\plotone{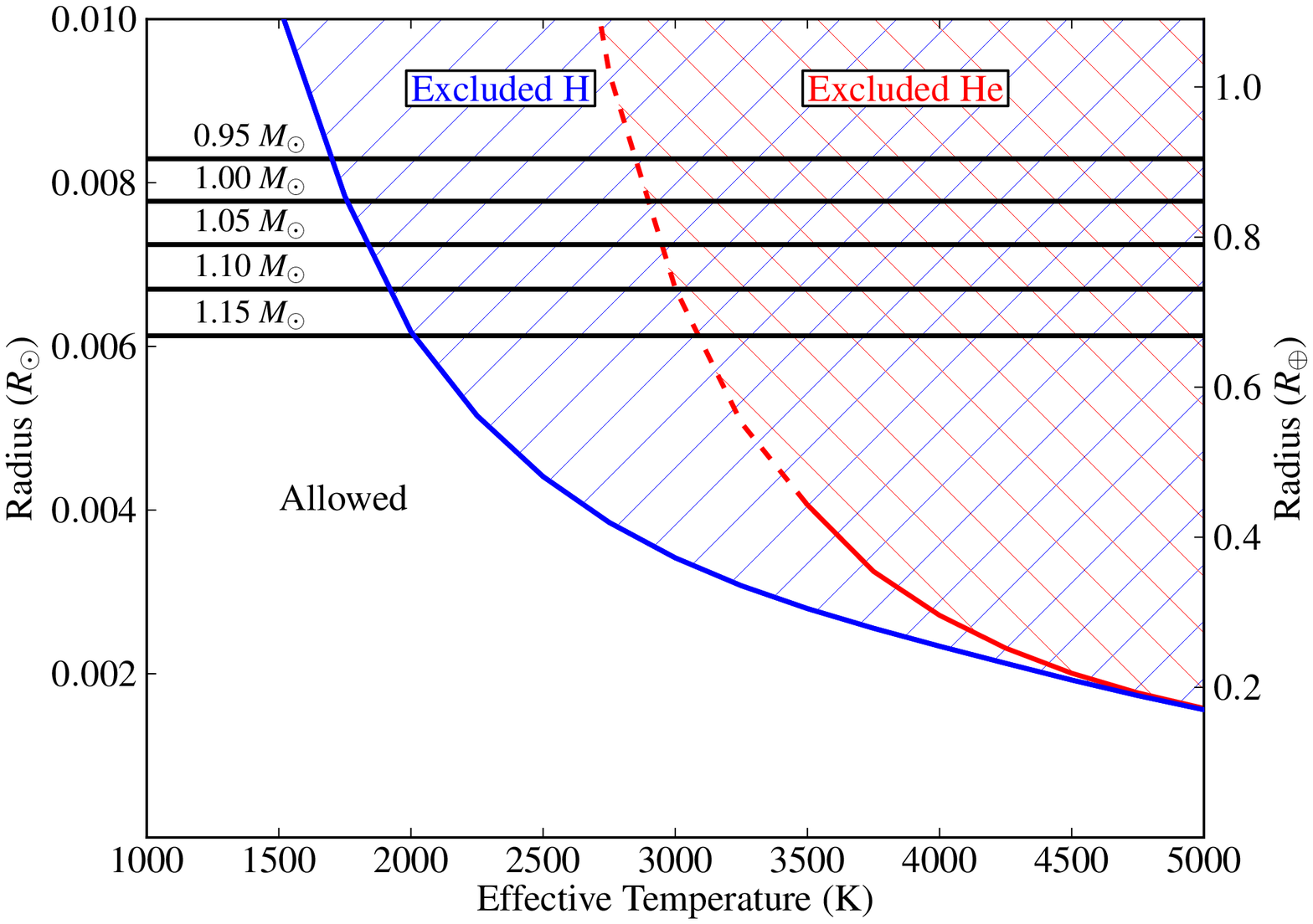}
\caption{Constraints on the radius of any WD companion to \psr, as a
  function of effective temperature.  The blue-hatched region shows
  the excluded parameter space for a H atmosphere (DA) white dwarf
  based on our $R$-band photometry (the $r$-band limit of $M_r>19.2$ is very
similar, given slight differences in bolometric correction), where we have used the synthetic
  model for the $1.0\,M_\odot$ DA WD to compute bolometric corrections
  as a function of effective temperature.  The red-hatched region
  shows the excluded parameter space for an He atmosphere (DB) white
  dwarf; we have extrapolated that model below 3500\,K with a
  blackbody (dashed segment), which is appropriate given the uncertainties in the
  equation-of-state in this regime.    The
  allowed radii should be compared with the radii of C/O WDs with masses from $0.95\,M_\odot$ to
  $1.15\,M_\odot$ (roughly our 2-$\sigma$ range from our mass
  measurements), shown by the horizontal lines.}
\label{fig:radius}
\end{figure}

The most constraining limit is again from the $R$-band data, where
$M_R>19.1$ implies $\Teff<1700\,$K (see \Fref{radius}) for a H
atmosphere.  The He-atmosphere models do not extend to sufficiently
cool temperatures but stop at $\Teff=3500\,$K with $M_R=17.7$.  At
lower temperatures the details of the atmospheric physics are rather
uncertain, but a blackbody is likely an acceptable approximation
(P.~Bergeron, 2014, private communication).  With a He atmosphere an effective
temperature $<3000\,$K would be required (\Fref{radius}).
The H limits are more constraining since more of the flux appears in
the optical regime rather than the near-infrared---a consequence of
collisionally induced absorption by molecular H$_2$
\citep*{bsw95,hansen98}.  These limits change slightly with mass given
the small but finite mass uncertainties, since the radius would change
with mass:
going to the $1.2\,M_\odot$ H model we can constrain $\Teff<2100\,$K
(at our nominal mass of $1.05\,M_\odot$ the radius of a C/O WD is about
$0.0073\,R_\odot$, and it scales as $R\propto
M^{-1.6}$).  As inferred from \Fref{mass}, the companion to
\psr\ would be far cooler
than any known WD from other surveys
\citep[e.g.,][]{ktka12,ctp+12,cnh+13}, where the coolest objects tend
to have $\Teff\approx 3800\,$K.

However, we cannot exclude such a very cool WD on age grounds.  WD
cooling curves, which start out having more massive objects warmer at
the same age, eventually cross to have more massive objects cooler at
the same age (\Fref{cool}; this is also visible in \Fref{mass}).  This
is because massive WDs crystallize earlier, at a higher \Teff\ (but at
a similar internal temperature), at which point the faster Debye
cooling takes over \citep{mr67,vanhorn68,cbfs00}.  Cooling ages for these models
may not be reliable, as the impacts of state changes, sedimentation,
and chemical processes are not precisely known, and the atmospheres
are not trivial to calculate \citep{mkww99,cbfs00,agi+07,scp+10,tbg11}.  But
we believe conservatively that the cooling age is close to 10\,Gyr,
almost certainly $>8\,$Gyr. In \Fref{cool} we show example
cooling curves, computed for thin and thick DA atmospheres
and C/O WDs (likely irradiation is a negligible perturbation to the
WDs surface temperature, given the measured spin-down luminosity of
the pulsar). For the model closest to the best-fit mass of \psr\ we
would infer that the true age is near 9\,Gyr, with the possible range
from 6--12\,Gyr.  The upper limit provided by the pulsar's
characteristic spin-down age (34\,Gyr after correction for the
Shklovskii effect [\citealt{shklovskii70}]) is not constraining; the assumption
that the pulsar's initial spin period is much shorter than the current
spin period is clearly not valid.  Instead, we take as our upper limit
to the age that of the Milky Way's halo ($11.4\pm0.7\,$Gyr;
\citealt{kalirai12}) minus the $\approx 70\,$Myr required for the
main-sequence lifetime of a $\approx 6\,M_\odot$ progenitor
\citep*{kr96,wbk09}, although this does not really exclude any models.  Such
an age would, however, imply a lower limit to the (re-)birth period of
about 25\,ms, assuming spin-down with a braking index $n=3$ (magnetic
dipole radiation).  We note
that the cooling models in \Fref{cool} may not be the only solution
for this progenitor: changing the WD composition (likely it is below
the transition to O/Ne/Mg WDs based on \citealt{nomoto84,it85},
although binary evolution could change that; also see
\citealt{ltk+14}) or atmosphere (helium, carbon, etc) could lead to
different solutions, and to draw robust conclusions we need to explore
a wider range of models with better observational constraints.  There are also considerable complications
and uncertainties in models for these temperatures: for instance, the
models of \citet[][the BaSTI database]{scp+10} give rather different
ages as \Teff\ never drops below $4000\,$K for $1.0\,M_\odot$ models,
even for ages of $>14\,$Gyr, while \citet{agi+07} and \citet{cbfs00}
do have $\approx 1.0\,M_\odot$ models go below 4000\,K (note that the
models in \citealt{agi+07} are primarily O/Ne rather than C/O).  However, we
believe the \Teff\ upper limits to be more robust, as they do tend to
agree between different calculations.

{
While extreme, the companion to \psr\ may not be especially unique.
Similar ultra-cool WDs are presumably present in globular clusters
and in the field even if they are often too faint to identify on
their own.  Individual ultra-cool WDs can be identified but only if very nearby,
like the two objects in \citet{ktka12}  at $\approx 30\,$pc.
If we correct roughly for the different progenitor masses between the
\citet{ktka12} systems and \psr\ 
\citep{khk+08} and use a \citet{salpeter55} initial mass function, we
would estimate $\approx 200$ massive WDs of a similar age within
300\,pc, which is of the same order as the luminosity function from
\citet[][also see \citealt{gbd12}]{rowell13} extrapolated\footnote{Similarly, \citet{gbd12} have a total WD number
  density of $\expnt{4}{-3}\,{\rm pc}^{-3}$ and \citet{rowell13} say that at most a few percent of WDs
are lost off the faint end of the luminosity function.} to $M_{\rm
  bol}>19$.}

{Instead, binary systems are the best way to identify cold WDs
  \citep[e.g.,][]{pgm+12}, which is effectively the technique used
  here.  But even in binary systems where we know that a source is
  present, the systems will often be too distant for good constraints
  (i.e., PSR~J1454$-$5846 in \Fref{mass}; \citealt{jcvk+06}).  We
  still require a fortuitously nearby system for useful observations.
  The
  occurrence of a nearby massive WD like the companion to \psr\ is reasonably
  consistent with expectations based on the observed binary
  population: there are five pulsar binaries from the ATNF Pulsar
  Catalog\footnote{\url{http://www.atnf.csiro.au/research/pulsar/psrcat/}.}
  \citep{mhth05} within 300\,pc, and the other four have low-mass He
  WD companions.  This $1/5$ ratio is similar to that for CO WD
  compared to He WD companions in the whole ATNF catalog (also see
  \citealt{tlk12}), and the pulsars' spin-down ages appear to have
  similar distributions for both companion types.}  

{Finally, we can ask whether  an NS is the most likely
  companion to an ultra-cool WD.  Most binaries are assumed to have mass
ratios near one (\citealt{ps06}, but see \citealt{sdmdk+12}), but a
binary composed of two ultra-cool WDs would be just as hard to detect
optically as a single object. If the companion were a lower-mass WD or
a main-sequence star the binary could be visible, although it would
require spectroscopic follow-up to identify the companion and in the
absence of \textit{GAIA} this has not been done for the majority of
stars within a few hundred pc.  So the situation of \psr, with an NS companion, is reasonably plausible as the initial mass
ratio would have been close to one and the chances of companion
follow-up and identification after discovery of the pulsar are high.}

\begin{figure}
\plotone{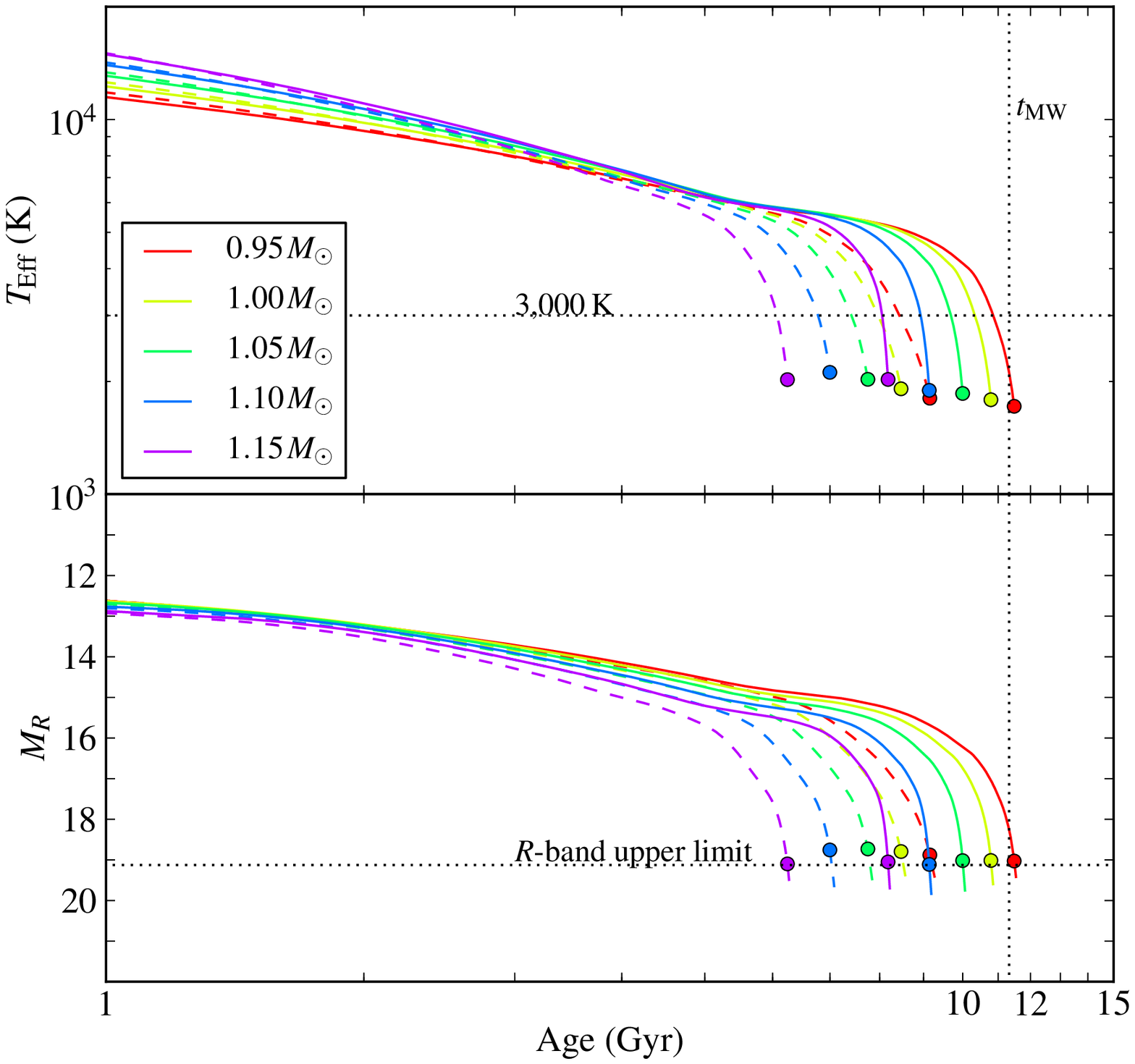}
\caption{Cooling of massive H-atmosphere CO WDs, based on the models
  of \citet[][also see \citealt{ks06,hb06,tbg11}]{bwd+11}.  We show
  the effective temperature (top), and $R$-band 
  (bottom) absolute magnitudes for ages of $>1\,$Gyr.  The models span
  the $\pm 2\sigma$ mass range of \psr's companion, from $0.95\,M_\odot$ to
  $1.15\,M_\odot$, and include thin (hydrogen $10^{-10}$ by mass;
  dashed lines) and thick (hydrogen $10^{-4}$ by mass; solid lines)
  hydrogen atmospheres.  The spin-down age of the pulsar $\tau_c$ is
  $34\,$Gyr and is far to the right.  Instead we show with a vertical
  dotted line the age of the Milky Way's inner halo \citep{kalirai12}
  as an upper bound to the age of any star not found in a globular
  cluster.  The $R$-band upper limits are the horizontal
  dotted lines in the lower panel: in the top panel the cooling curves stop when the
  implied $R$-band photometry reaches our upper limit (filled circles), which happens at an
  effective temperature of 2,000--3,000\,K (3,000\,K is indicated by
  the dotted line in the top panel). To compute the
  synthetic photometry we have used the synthetic model for the
  $1.0\,M_\odot$ DA WD to compute bolometric corrections as a function
  of effective temperature, which we then applied to the cooling
  models.  }
\label{fig:cool}
\end{figure}

\section{Conclusions}
\label{sec:conc}
We have determined an accurate mass for the partially recycled pulsar
\psr\ and its companion; the latter is  value consistent with both an NS and a WD.  Despite not finding the companion in
a deep optical/near-infrared search, we reject a DNS
explanation as the binary system shows evidence of circularization
requiring mass transfer after the last supernova.  Instead the
companion is likely a high-mass WD.  Using the extremely
precise distance determination from \citet{dbl+13}, we can set a
robust limit of $M_R > 19.1$.  This implies an very old and cool
WD: fainter than all other pulsar companions by a factor of
about 100, and fainter than the lower-mass ``ultra-cool'' WD
in the solar neighborhood by a factor of about four.  Converting this
limit to a temperature depends somewhat on the assumed mass and
composition, but we believe an effective temperature limit of
$\Teff<3000\,$K is a robust upper limit. For such an object to not be
older than the Milky Way requires that it have already entered the
faster Debye cooling regime, i.e., that it already crystallized (also
see \citealt*{mmk04,bf05}).  Future searches, if they can detect the
companion to \psr, will be a unique probe of the very late stages of
WD evolution, with a well-determined mass and radius that are
not usually available for studies of such objects.

\acknowledgements We thank an anonymous referee for useful
suggestions, and T.~Tauris, M.~van~Kerkwijk, I.~Stairs, P.~Bergeron and
R.~O'Shaughnessy for helpful discussions.  D.L.K.\ is supported by the
National Science Foundation grant AST-1312822. M.A.M.\ and D.R.L.\ are
supported by WVEPSCOR, the NSF PIRE Program, and the Research
Corporation for Scientific Advancement. JRB acknowledges support from
WVEPSCoR, the National Radio Astronomy Observatory, the National
Science Foundation (AST 0907967), and the Smithsonian Astrophysical
Observatory (Chandra Proposal 12400736).  A.T.D.\ was supported by an NWO
Veni Fellowship.  Some of the data presented herein were obtained at
the W.~M.\ Keck Observatory, which is operated as a scientific
partnership among the California Institute of Technology, the
University of California and the National Aeronautics and Space
Administration. The Observatory was made possible by the generous
financial support of the W.~M.\ Keck Foundation.  The authors wish to
recognize and acknowledge the very significant cultural role and
reverence that the summit of Mauna Kea has always had within the
indigenous Hawaiian community.  We are most fortunate to have the
opportunity to conduct observations from this mountain.  Based on
observations obtained at the Southern Astrophysical Research (SOAR)
telescope, which is a joint project of the Minist\'{e}rio da
Ci\^{e}ncia, Tecnologia, e Inova\c{c}\~{a}o (MCTI) da Rep\'{u}blica
Federativa do Brasil, the U.S. National Optical Astronomy Observatory
(NOAO), the University of North Carolina at Chapel Hill (UNC), and
Michigan State University (MSU). Funding for SDSS-III has been
provided by the Alfred P. Sloan Foundation, the Participating
Institutions, the National Science Foundation, and the U.S. Department
of Energy Office of Science. The SDSS-III Web site is
http://www.sdss3.org/.  We made extensive use of SIMBAD, ADS, and
Astropy (\url{http://www.astropy.org}; \citealt{astropy}).

{\it Facilities:}  \facility{GBT}, \facility{Keck:I (LRIS)}, \facility{Keck:II (NIRC2)}

\bibliographystyle{apj} 

\end{document}